%% file: paper.tex
\documentclass[sigconf]{acmart}

\AtBeginDocument{%
  \providecommand\BibTeX{{%
    \normalfont B\kern-0.5em{\scshape i\kern-0.25em b}\kern-0.8em\TeX}}}

\setcopyright{acmcopyright}
\copyrightyear{2022}
\acmYear{2022}
\acmDOI{10.1145/1122445.1122456}

\acmConference[ICSE '22]{ICSE '22: International Conference on Software Engineering}{May 21--29, 2022}{Pittsburgh, PA, USA}
\acmBooktitle{ICSE '22: International Conference on Software Engineering,
  May 21--29, 2022, Pittsburgh, PA, USA}
\acmPrice{15.00}
\acmISBN{978-1-4503-XXXX-X/18/06}

\usepackage{caption}
\usepackage{subcaption}
\usepackage{algorithmic}
\usepackage{graphicx}
\usepackage{textcomp}
\usepackage{booktabs}
\usepackage{xcolor}
\usepackage{xspace}
\usepackage{colortbl}
\usepackage{tabularx}
\usepackage{soul}

\usepackage{multirow, makecell}
\colorlet{lightgrey}{lightgray}

\usepackage{array}
\def\BibTeX{{\rm B\kern-.05em{\sc i\kern-.025em b}\kern-.08em
    T\kern-.1667em\lower.7ex\hbox{E}\kern-.125emX}}

\usepackage{tikz}
\usepackage{pgfplots}
\pgfplotsset{width=7.5cm,compat=1.12}
\usepgfplotslibrary{fillbetween}

\def \arc {\textsc{ARC}\xspace}
\def \de {\textsc{D}\xspace}
\def \arcl {\textsc{ARC\textsubscript{L}}\xspace}
\def \token #1{{\tt [#1]}}
\def \nl {\textsc{BR\textsubscript{NL}}\xspace}
\def \pl {\textsc{BR\textsubscript{PL}}\xspace}
\def \fl {\textsc{BR\textsubscript{FL}}\xspace}
\def \bert {FBL-BERT\xspace}

\definecolor{green1}{HTML}{004c00}
\definecolor{green2}{HTML}{009900}
\definecolor{green3}{HTML}{47da47}
\definecolor{green4}{HTML}{99ea99}
\definecolor{blue1}{HTML}{003566}
\definecolor{blue2}{HTML}{186fc0}
\definecolor{blue3}{HTML}{68aae7}
\definecolor{blue4}{HTML}{a1caf0}
\definecolor{red0}{HTML}{E74C4C}
\definecolor{red1}{HTML}{990000}
\definecolor{red2}{HTML}{DD0000}
\definecolor{red3}{HTML}{E74C4C}
\definecolor{red4}{HTML}{F19999}
\definecolor{yellow1}{HTML}{F1C239}
\definecolor{gray1}{HTML}{C8C8C8}
\definecolor{gray2}{HTML}{707070}
\definecolor{gray3}{HTML}{505050}
\definecolor{brown1}{HTML}{654321}

\usepackage{hyperref}
\usepackage{xparse}
\usepackage{etoolbox}

\newrobustcmd*{\mycircle}[1]{\tikz{\filldraw[draw=#1,fill=#1] (0,0) circle [radius=0cm] (0,0.1cm) circle [radius=0.065cm];}}

\newrobustcmd*{\mytriangle}[1]{\tikz{\filldraw[draw=#1,fill=#1] (0,0) circle [radius=0cm] (0,0.03cm) -- (0.12cm,0.03cm) -- (0.06cm,0.15cm);}}

\newrobustcmd*{\mysquare}[1]{\tikz{\filldraw[draw=#1,fill=#1] (0,0) circle [radius=0cm] (0.05cm,0.05cm) rectangle (0.15cm,0.15cm);}}

\begin{document}

\title{Fast Changeset-based Bug Localization with BERT}

\author{Agnieszka Ciborowska}
\affiliation{%
  \institution{Virginia Commonwealth University}
  \department{Department of Computer Science}
  \city{Richmond}
  \state{VA}
  \country{USA}}
\email{ciborowskaa@vcu.edu}

\author{Kostadin Damevski}
\affiliation{%
  \institution{Virginia Commonwealth University}
  \department{Department of Computer Science}
  \city{Richmond}
  \state{VA}
  \country{USA}}
\email{kdamevski@vcu.edu}

\renewcommand{\shortauthors}{Ciborowska and Damevski}

\begin{abstract}
\input{abstract}
\end{abstract}


\maketitle

\input{intro}

\input{problem}

\input{approach}

\input{evaluation}

\input{results}

\input{related}

\input{conclusion}

\bibliographystyle{ACM-Reference-Format}
\bibliography{paper}

\end{document}

%% file: abstract.tex
Automatically localizing software bugs to the changesets that induced them has the potential to improve software developer efficiency and to positively affect software quality. To facilitate this automation, a bug report has to be effectively matched with source code changes, even when a significant lexical gap exists between natural language used to describe the bug and identifier naming practices used by developers. To bridge this gap, we need techniques that are able to capture software engineering-specific and project-specific semantics in order to detect relatedness between the two types of documents that goes beyond exact term matching. Popular transformer-based deep learning architectures, such as BERT, excel at leveraging contextual information, hence appear to be a suitable candidate for the task. However, BERT-like models are computationally expensive, which precludes them from being used in an environment where response time is important.

In this paper, we describe how BERT can be made fast enough to be applicable to changeset-based bug localization. We also explore several design decisions in using BERT for this purpose, including how best to encode changesets and how to match bug reports to individual changes for improved accuracy. We compare the accuracy and performance of our model to a non-contextual baseline (i.e., vector space model) and BERT-based architectures previously used in software engineering. Our evaluation results demonstrate advantages in using the proposed BERT model compared to the baselines, especially for bug reports that lack any hints about related code elements.

%% file: intro.tex
\section{Introduction}

Two of the most prevalent tools used today by software engineers are repositories to store project files (e.g., git) and bug trackers to report and monitor bug fixing activity (e.g., JIRA, BugZilla). Automatically linking a bug report in a bug tracker and related software artifacts from a repository is one of the long-standing goals in the software engineering research community, due to its potential to improve practice by reducing the time developers spend examining code when addressing a newly reported bug, i.e., {\em bug localization}~\cite{bugscout,zhou2012where}. However, despite numerous efforts, the accuracy of bug localization approaches is not yet high enough for widespread use, especially as it applies to different software projects that vary in bug report and code style~\cite{mills2020relationship}. In examining the trends from interviews conducted with a large cohort of software developers from industry and open-source software, Zou et al. report that developers do not trust bug localization tools due to their inability to adapt to different types of bug reports, specifically noting that existing techniques only work on the most simple cases, with straightforward textual similarity between the bug report and code base~\cite{zou2020how}. More work is needed to improve the retrieval quality of bug localization techniques. 

At the same time, as industry is increasingly attempting to use bug localization to aid developers in their daily work, specific requirements of the problem for modern use are coming to the forefront~\cite{murali2020industry}. One key characteristic found beneficial in modern software projects is bug-inducing changeset- (or commit-) level retrieval. A bug-inducing changeset is one where the bug was initially introduced into the repository. Retrieving such changesets leads to faster bug repair, as they contain related parts of the code that were changed together, which makes fixing the bug easier. However, retrieving bug-inducing changesets with high accuracy is more challenging than retrieving buggy source code elements due to the potentially large number of commits in the corpus.

In recent years, numerous popular natural language processing tasks (e.g., question answering, machine translation) have all observed improved performance when using neural network architectures based on transformers. 
These transformer-based models are typically applied via transfer learning, by first pre-training them on a very large corpus and then fine tuning on a much smaller dataset towards the specific task they are to be used for.
Transformer-based models pre-trained on large software engineering corpora (e.g., StackOverflow) are now becoming available~\cite{tabassum-etal-2020-code}, with the potential to improve software engineering tasks like bug localization. In this paper, we use the BERT (Bidirectional Encoder Representations from
Transformers) transformer-based architecture, which is a highly popular model introduced by Devlin et al.~\cite{devlin2019bert}. 

Bug localization is usually framed as an Information Retrieval (IR) task, where a document (i.e., a software artifact) is retrieved from a corpus-based on a query (i.e., the bug report text). 
A measure of semantic relatedness between the bug report and the software artifact is necessary to rank the results retrieved from the corpus. Given the fact that transformer-based models consist of many neural layers and require heavy computation for each sentence, measuring relatedness between the query and the corpus quickly becomes expensive.

This paper applies BERT to the problem of changeset-based bug localization with the goal of improved retrieval quality, especially on bug reports where straightforward textual similarity would not suffice. We describe an architecture for IR that leverages BERT without compromising retrieval speed and response time. In addition, we examine a number of design decisions that can be beneficial in leveraging BERT-like models for bug localization, including how best to encode changesets and their unique structure.

Our experimental results indicate that the proposed approach improves upon popular bug localization techniques by, e.g., increasing the retrieval accuracy between 5.5\% and 20.6\% for bug reports with no or a limited number of localization hints. We note that using entire changesets as input granularity significantly hinders the models performance, while leveraging more fine grained input data, such as hunks, results in the highest retrieval quality. We also observe that the size of search space (i.e., the number of changesets in a project) significantly impacts the retrieval delay of different BERT-based models, though less in the case of the proposed model. 

The main contributions of this paper are:
\begin{itemize}
    \item \textbf{approach that applies BERT to the bug localization problem} (specifically, localizing bug-inducing changesets) that is more accurate than the state-of-the-art,
    \item \textbf{improvement over other recent BERT-based architectures} proposed towards changeset retrieval, showing significant advantages with respect to retrieval speed,
    \item \textbf{evaluation and recommendations for key design choices} in applying BERT to changesets (i.e., code change encoding, data granularity).
\end{itemize}

\noindent
\textbf{Significance of contribution.} The BERT-based technique proposed in this paper enables semantic retrieval of software artifacts (specifically, changesets) for bug localization that goes beyond (and can complement) the exact term matching in the current popular state-of-the-art techniques (e.g.,~\cite{saha_bluir_2013,wong_brtracer_2014}). Relative to a similar, recent BERT-based technique~\cite{lin2021traceability}, we offer an approach that improves retrieval speed significantly, in a way that supports real-world use, while also enhancing retrieval quality.   

%% file: problem.tex
\section{Problem Description}

In this section, we list and discuss the specific constraints of the bug localization problem that we aim to address, which are based on a recent survey of industry practitioners and the problem requirements observed at a large software enterprise~\cite{murali2020industry,zou2020how}. Our focus is a bug localization technique that: 1) focuses on retrieving changesets; 2) aims to capture semantics and can be applied to bug reports that do not share terms with the relevant parts of the code base; and 3) quickly retrieves results for a newly created bug report.

\smallskip
\noindent
\textbf{1. Localizing changesets}~\cite{murali2020industry}.
Over the years, a large body of research has been dedicated to locating source code files (or classes) relevant to a bug report~\cite{kim_buglocator_2013,saha_bluir_2013,wong_brtracer_2014,wang_amalgam_2014,bug_scout,corley2018}. However, recent studies have pointed out that bug localization at the level of source code files still requires significant effort by software developers in order to locate relevant code within large files~\cite{wen2016, murali2020industry, zou2020how}. Adjusting for this finding, researchers shifted their efforts towards more fine grained code elements, such as file segments~\cite{wong_brtracer_2014} and methods~\cite{ye2014learning, tantithamthavorn2018impact,zhang2019finelocator}, which introduce new sets of challenges such as difficulty in selecting optimal segment size and large methods that still require effort to examine. More recently, there has been a growing interest in changeset retrieval~\cite{corley2018, wen2016, lin2021traceability, wu2018change} for bug localization because changesets have several unique properties that make them convenient to developers aiming to fix a bug. First, they inherently capture lines of code that are related to each other within the context of a modification. 
Second, when locating changesets, we can retrieve not only the modified portion of the code, but identify a software developer that committed the modification in the first place, therefore easing the bug triaging process. Finally, changesets allow for straightforward context-aware division into a set of hunks, i.e., a set of changes in one area of the file. Hunks are usually convenient to read for developers and allow for easy detection of changes with no semantic value (e.g., changes only in white spaces).

\smallskip
\noindent
\textbf{2. Leveraging semantics of input documents}~\cite{murali2020industry,tabassum-etal-2020-code}. 
As software evolves rapidly and is actively maintained by multiple developers, different portions of the code base become affected by distinctive identifier naming patterns and conventions, which exacerbate the already existing semantic gap between bug reports and related code elements, posing a significant challenge to traditional IR systems based solely on token similarity~\cite{guo2017semantically}. Surveys of practitioners have also indicated that bug reports that explicitly mention the names of classes or methods relevant to the bug fix do not require automated bug localization, while assisting in bug reports with large semantic gaps with the code base is likely more valuable to developers. For instance, one surveyed developer in the study by Zou et al.~\cite{tabassum-etal-2020-code} stated the following about current bug localization, {\em "It seems that existing techniques mainly make use of the textual
similarity between bug reports and source code files to
perform bug localization. However, I encountered many
bugs that have very little similarity between their bug
reports and code files. I wonder what kind of bugs such
techniques can localize? Maybe only simple bugs?"}.  To bridge this gap, researchers have recently proposed to use deep learning models capable of building semantically rich document representations~\cite{guo2017semantically, lin2021traceability,lam2015,huo2019deep,cheng2020similarity,murali2020industry,cao2020bug}. 
Transformer-based models, and BERT in particular, are currently one of the most exciting deep learning techniques achieving broad improvements across a variety of text-based tasks. 
The main strength of  BERT-like models is in building a token representation based on bidirectional contextual information encoded in the preceding and succeeding tokens, which leads to richer semantics that is more likely to detect related pairs of bug reports and changesets that do not share terms. Prior generations of word embeddings, e.g., word2vec~\cite{mikolov2013distributed} and GloVe~\cite{pennington2014glove}, which have been frequently applied on software engineering tasks~\cite{chen2019literature}, do not use word context at inference time, i.e., each token maps to a
vector regardless of the surrounding text.

\begin{figure*}[th]
     \centering
     \begin{subfigure}[b]{0.28\textwidth}
         \centering
         \includegraphics[width=\textwidth]{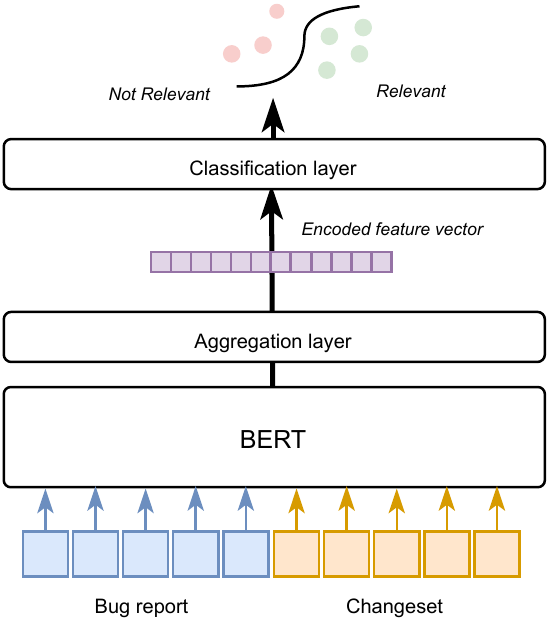}
         \caption{Single BERT~\cite{lin2021traceability, dai2019deeper,nogueira2020passage}}
         \label{fig:single-bert}
     \end{subfigure}
     \hfill
     \begin{subfigure}[b]{0.28\textwidth}
         \centering
         \includegraphics[width=\textwidth]{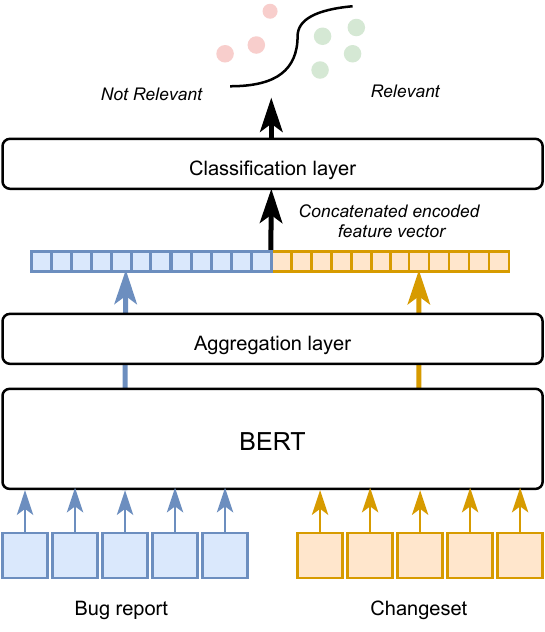}
         \caption{Siamese BERT~\cite{lin2021traceability, reimers2019sentencebert}}
         \label{fig:siamese-bert}
     \end{subfigure}
     \hfill
     \begin{subfigure}[b]{0.28\textwidth}
         \centering
         \includegraphics[width=\textwidth]{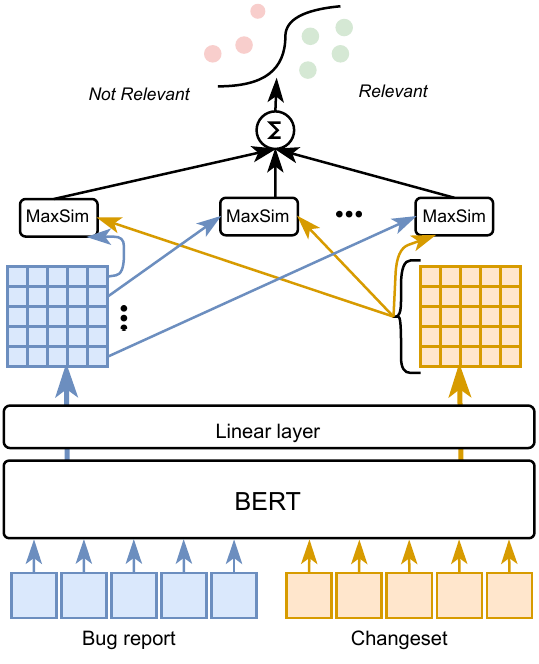}
         \caption{FBL-BERT~\cite{khattab2020colbert}}
         \label{fig:colbert}
     \end{subfigure}
        \caption{BERT-based architectures for changesets retrieval}
        \label{fig:berts}
\end{figure*}

\smallskip
\noindent
\textbf{3. Fast retrieval in a large search space}~\cite{pradel2020}. Retrieving bug-inducing changesets requires computing similarity between a bug report of interest and all changesets committed to a repository up to the present point in time. Given that modern software evolves rapidly, resulting in large source code repositories with numerous commits~\cite{savor_facebook,commitguru}, it is impractical to compute pair-wise similarity due to the large search space. This is especially the case if computing the similarity measure itself is expensive. 
Though deep learning models provide state-of-the-art accuracy, they typically require more computational resources than token-based techniques, which emphasizes the need for a bug localization technique to limit the search space in order to improve performance without compromising accuracy.

%% file: approach.tex
\section{Approach}

In order to address the above problem constraints, in this paper we investigate the use of a BERT model towards bug localization with changesets as a primary data granularity, which is also preferred by practitioners' (\S 2.1)\cite{wan2018perceptions}. We specifically selected BERT as it is the state-of-the-art in semantics modeling and extracting contextual information (\S 2.2). 
Finally, to ensure that our approach is applicable to large, industry scale repositories (\S 2.3), we introduce Fast Bug Localization BERT (FBL-BERT), which reduces the search space, such that only promising candidate changesets are considered for neural re-ranking with BERT. In addition, \bert encodes a bug report and a changeset separately, allowing to compute changeset representations offline and reduce the computational effort per bug report at retrieval time. Replication package is available at~\cite{repo}.

\setlength{\belowcaptionskip}{5pt}
\begin{figure*}[ht]
    \centering
    \includegraphics{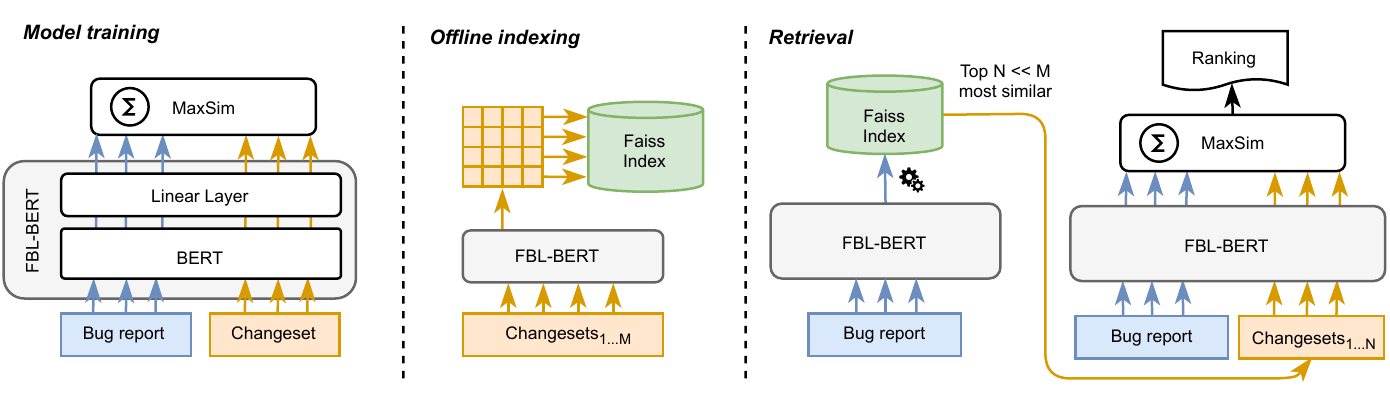}
    \caption{\bert for changeset-based bug localization pipeline.}
    \label{fig:pipline}
\end{figure*}

\subsection{BERT for bug localization}

The architecture of BERT consists of multiple layers of transformer-encoders, which are an abstraction aimed at modeling sequential data that utilizes self-attention; the notion of attention is to weight specific terms in the sequence differently, i.e., encoding a stronger relationship from each term in the sequence to the remaining most semantically relevant terms. 
As pointed out by Mills et al.~\cite{mills2020relationship}, retrieval techniques for bug localization can be significantly improved with intelligent query construction, i.e., by carefully choosing which parts of the bug report to use for comparison. Therefore, leveraging a model that uses attention to emphasize certain word relationships has the potential to significantly improve upon prior state-of-the-art bug localization techniques.

Using a BERT model for bug localization (or other similar purposes) involves three essential steps: (1) pre-training the model with a large corpus of general software engineering-related data, (2) fine tuning the BERT model for bug localization, and finally, after BERT has been completely trained, (3) retrieving relevant bug-inducing changesets for a newly reported bug. 

During pre-training, BERT uses massive corpora of relevant text to build a  language model for a specific domain, e.g., software development.
Given that this step requires a significant amount of data and computational resources, a common choice is to re-use a pre-trained BERT model, when available.
In the fine tuning step, BERT updates the general data representation with respect to a specific downstream task (e.g., bug localization) given a much smaller, task-specific dataset. More precisely, fine tuning a BERT model occurs by adding an additional layer (e.g., a classification layer) to the pre-trained BERT model. This task-specific layer takes the output of BERT as input and represents the part of the model that is primarily trained during fine tuning, though BERT's internal weights are also updated in the process. In most scenarios, fine tuning can be completed faster and with much less computational resources than pre-training. Since our goal is locating bug-inducing changesets, a natural choice for a task-specific dataset consists of bug reports and their inducing changesets.
A key design choice at this stage is how to connect BERT with the additional task-specific neural network layer. 
Given an input document, BERT encodes each word in the document with a vector, i.e., for each input document, the output of the BERT model is an embedding matrix of size $|d|$ by $v_{len}$, where $|d|$ represents the number of words in the document and $v_{len}$ the length of a BERT vector; typically $v_{len}=728$. 
The most common approach when retrieving BERT-encoded documents is to aggregate the embedding matrix across words through average or summation, which produces a single vector as output. Using such an aggregate representation of a document allows for faster processing and easier comparison between pairs of documents. However, as pointed by Sachdev et al.~\cite{sachdev2018retrieval}, this simple aggregation strategy leads to a dissipative data representation that has the potential to negatively affect retrieval performance. In the next section, we describe an alternative strategy that takes advantage of the full matrix to encode input data. 

In the simplest changeset retrieval scenario, presented in Fig.~\ref{fig:single-bert}, each newly arriving bug report is concatenated with every changeset in the project history. Subsequently, they are processed by BERT, producing an embedding matrix, which is transformed to a vector by an aggregation layer. Finally, the vector is passed into a classification layer that produces a relevancy score between a bug report and a changeset. Changesets are ordered based on their scores to produce a ranked result set. This type of BERT architecture for information retrieval is often referred to as Single BERT~\cite{lin2021traceability, dai2019deeper,nogueira2020passage}. In an alternative retrieval architecture, called Siamese BERT~\cite{lin2021traceability, reimers2019sentencebert} and depicted in Fig.~\ref{fig:siamese-bert}, the bug report and the changeset are processed separately, first through BERT and then through an aggregation layer. As a result, bug reports and changesets are transformed into independent vectors that are subsequently concatenated and fed into the classification layer to produce a relevance score. 
The advantage of Siamese BERT over Single BERT is that Siamese BERT enables pre-computing changeset representations offline since changesets are not required to be concatenated with a bug report for retrieval. However, Siamese BERT still requires comparing a bug report to each changeset, which incurs significant retrieval delay in the case of large number of changesets.

\subsection{Fast Bug Localization BERT}
\label{sec:late-interaction}
The \bert architecture, based on ColBERT by Khattab et al.~\cite{khattab2020colbert}, eschews aggregation of the embedding matrix, and instead builds a relevance score by leveraging the whole matrix, resulting in a more complete, fine grained comparison. More specifically, a bug report $br$ and a changeset $c$ are separately processed by BERT creating embedding matrices $E_{br}$ and $E_c$, respectively. To compute the relevance score between $E_{br}$ and $E_{c}$, for each word embedding in the bug report $v_{br} \in E_{br}$, we find the maximum cosine similarity across word embeddings of the changeset $v_c \in E_c$, and combine the maximum cosine similarities via summation as illustrated in Fig.~\ref{fig:colbert}. 
As a result, the model learns how to associate words from a bug report with tokens in a changeset, taking into account the context in which they appear. 
To account for the two different types of data we process, i.e., bug reports and changesets, we modify ColBERT by increasing the numbers of BERT encoder layers taken to the linear layer. More specifically, while ColBERT uses the output of the last BERT encoder, we take the output of the last 4 encoders (as recommended by~\cite{devlin2019bert}). This modification is dictated by prior studies observing that that different layers of BERT encode different granularity of semantic information~\cite{devlin2019bert, tenney2019bert, peters2018dissecting}.
Note that the linear layer in \bert is not equivalent to the aggregation layer discussed before, but is used to reduce the size of word embeddings produced by BERT, retaining all word embeddings in a compressed form for faster downstream processing. 

There are several benefits that make the \bert architecture particularly applicable to our problem. First, the model purposely avoids joint document encoding, as in Single BERT, delaying interaction between a bug report and a changeset to facilitate off-line encoding of changesets. 
Moreover, by using computationally cheap, yet efficient, maximum similarity summation as a scoring operator instead of a more complex strategy, such as the classification layer in Siamese BERT, the processing time for a query is reduced. 
Finally, given that the relevance score computation is isolated and relies solely on maximum similarity, it is possible to utilize efficient vector similarity algorithms to reduce the search space of all $M$ changesets by identifying top-$N$ changesets, $N<<M$, that are similar to a new bug report, and subsequently re-rank only the top-$N$ subset.

To clarify how \bert operates for changeset-based bug localization, consider the pipeline depicted in Fig.~\ref{fig:pipline}. First, as shown in the Model Training section of Fig.~\ref{fig:pipline}, the \bert model is fine tuned on a project-specific dataset consisting of bug reports and bug-inducing changesets.
In the next step (Offline Indexing), \textit{all changesets} in the project repository are encoded via \bert and stored in an index supporting efficient vector-similarity search. For this purpose, we use an IVFPQ  (InVerted File with Product Quantization) index, implemented in the Faiss library~\cite{JDH17}. The IVFPQ index uses the k-means algorithm to partition the embedding space into $P$ (e.g., $P=300$) partitions, and subsequently assigns each word embedding to its nearest cluster. To facilitate efficient search, when a query is issued, the query is first compared against the partitions' centroids to locate the nearest partitions, and then the search continues to the instance-level only within those.
Note that the Faiss index contains \textit{word} embeddings across \textit{all} changesets. After completion of this step, the retrieval system is ready to be deployed. When a new bug report arrives, it is first encoded via \bert producing an embedding matrix. Next, for each word embedding in the embedding matrix,  we query the Faiss index to identify the $N'$ most similar embeddings across all changesets embeddings stored in the Faiss index. Since among $N'$ most similar embeddings some may point to the same changeset, in the end we obtain a total of $N$ unique candidate changesets. Finally, we use \bert to re-rank the candidate changesets and produce the final ranking.

\subsection{Changesets encoding strategies}

Software evolution over time is recorded in a repository as a time-ordered sequence of changesets. Each changeset consists of a log message, providing a short rationale explaining the goal of the modification, and a set of source code changes. Depending on the version control system and \texttt{diff} algorithm used in the software project, the representation of source code changes can vary. In this paper, we focus on the format that is the output of the \texttt{git diff} command, in which added lines of code are annotated with \texttt{+}, removed lines with \texttt{-}, and all modified lines are surrounded by 3 lines of contextual, unchanged lines. While there exist more advanced tree-based code differencing algorithms (e.g., GumTreeDiff~\cite{falleri2019fine}), providing detailed code-change information to a machine learning model may affect the model negatively~\cite{zeng2021deep}, hence we opt for a text-based approach. Changesets can encapsulate code changes across one or multiple source code files, and modifications to each file can be divided into {\em hunks} - groups of modified (added or removed) lines surrounded by unchanged (context) lines. Given this specific formatting, we explore how best to utilize changesets' properties to construct BERT input from two perspectives: (1) encoding characteristics of code modifications, such as additions or removals; and (2) levels of granularity in a changeset.

\begin{figure}[ht]
    \centering
    \includegraphics[scale=1.15]{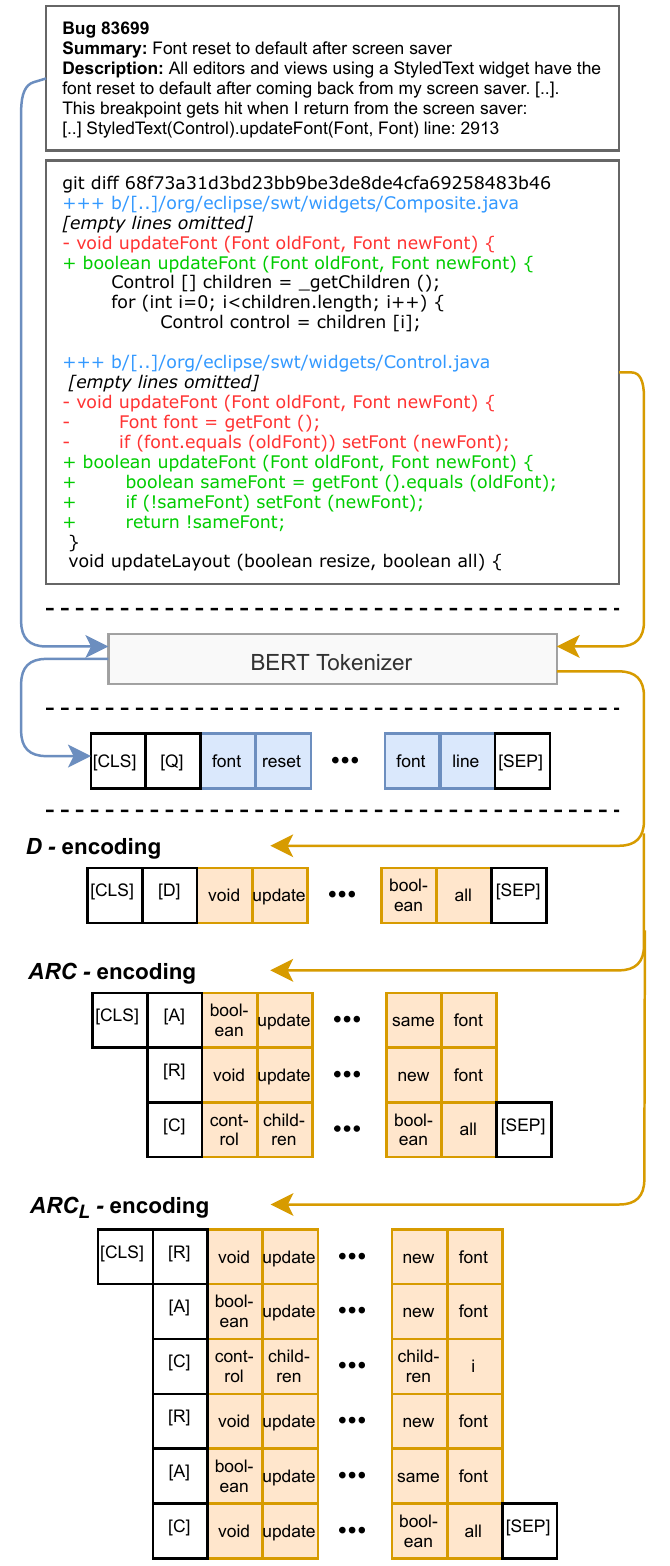}
    \caption{Changeset encoding strategies.}
    \label{fig:encoding}
\end{figure}

Input provided to BERT models is required to follow certain rules. First, a document (e.g., a changeset or a bug report) needs to be tokenized and each token replaced by its unique token id. Pre-trained BERT models supply their own BERT tokenizers, that are optimized towards the corpus on which the model is pre-trained. BERT tokenizers are trained using the WordPiece algorithm~\cite{schuster2012japanese}. The main advantage of BERT tokenizers is in avoiding out-of-vocabulary words by dividing unknown words to their largest subwords present in the vocabulary, which is likely to be beneficial in our setting, as software projects can have very specific vocabularies unlikely to be observed elsewhere~\cite{tabassum-etal-2020-code}.
Secondly, BERT uses a pre-defined set of special tokens. In general, due to how BERT is trained (more details in \cite{devlin2019bert}), the model requires that each token sequence starts with special classification token \token{CLS} and ends with separator token \token{SEP}, while other special tokens, such as padding \token{PAD} are used if and when necessary. Special tokens can convey information about the structure of data allowing BERT to differentiate between parts of the input, hence we explore how special tokens can be best utilized to encode changesets.
To this end, we propose the following encoding strategies, depicted in Fig.~\ref{fig:encoding}.

\smallskip
\noindent
\textbf{\de:} A changeset is considered a single document that is feed into the model. To inform the model that a changeset sequence begins, we define and pre-append the special token \token{D} at the beginning of the code sequence.
Since this strategy does not utilize specific characteristics of a code change, it serves as a baseline to compare against other strategies.

\smallskip
\noindent
\textbf{\arc:} In this encoding, a changeset is split into lines, and the lines are subsequently grouped based on whether they are added, removed or provide context, as indicated by their initial character: \texttt{+} for added, \texttt{-} for removed, and an empty space for context lines. The lines in each group are concatenated to create a sequence to which we pre-append a special token: \token{A} for the sequence of added lines, \token{R} for the sequence of removed lines and \token{C} for the sequence of context lines. Finally, all the sequences are concatenated together to create an input for the model.
By grouping different parts of changesets based on their characteristics, we aim to investigate whether any particular type of modification is more beneficial than the other. With the \arc strategy the model is given an opportunity to learn how to combine information of different types and, if necessary, decide to disregard a portion of it if it poorly affects performance.

\smallskip
\noindent
\textbf{\arcl:} Similarly as in \arc, a changeset is divided into lines, however \arcl encoding does not group the lines. Instead, it preserves the ordering of lines within a changeset, such that special tokens \token{A}, \token{R}, or \token{C} are pre-appended wherever type of modification changes. While this strategy results in more accurate data representation, compared to \arc, \arcl is also more challenging for the model, since the special tokens occur multiple times and in several places.
\noindent

\smallskip
Given that a bug report and a changeset are encoded separately, the model has to differentiate between these two types of documents. To this end, when encoding a bug report, we define a special token \token{Q} that is pre-appended to the query, i.e., the bug report.

Another dimension in choosing how to best encode changesets is related to their granularity, i.e., using entire changesets or separating a changeset to a file- or hunk-level.
Leveraging hunks as the primary data dimension in an IR model brings several advantages. First, bugs have been observed to be typically caused by small pieces of code~\cite{ye2014learning, wong_brtracer_2014}, thus the inherent fine granularity of hunks makes them less susceptible to noise when compared to whole source code files~\cite{wen2016}. Second, dividing changesets into hunks alleviates issues caused by tangled commits~\cite{kim2013impact}. Given the fact that hunks are typically small and concentrate on an enclosed portion of the code, BERT is not affected by long-range token dependencies, which is a problem typically affecting source code~\cite{tabassum-etal-2020-code}. Finally, shorter input documents are less likely to exceed the maximum sequence length accepted by BERT, while longer documents have to be truncated, which may negatively affect the results. However, despite easily accessible smaller data granularity within a changeset, to date, most of the efforts are focused on leveraging entire changesets~\cite{lin2021traceability,bhagwan2018orca,murali2020industry}.

%% file: evaluation.tex
\section{Experimental evaluation}

\subsection{Research questions}
\noindent
\textbf{RQ1:} \textit{How effective is \bert when compared to (1) state-of-the-art techniques based on the VSM, and (2) related BERT-based architectures?}

\noindent
The main opportunity in using \bert is in incorporating additional context and semantics when retrieving bug-inducing changesets, which should provide improvements in accuracy over the state-of-the-art, especially for bug reports that provide high level bug descriptions and lack explicit localization hints. Researchers have identified that a non-trivial amount of bug reports already contain localization hints, i.e., they
mention the class or method names relevant to fixing the bug, and some recent approaches for bug localization argue that only bug reports that lack extensive localization hints should be considered in evaluation~\cite{deeptransfer}.
We follow the methodology proposed by Kochhar et al.~\cite{kochhar2014} to categorize bug reports into 3 groups based on the completeness of localization hints they provide and evaluate the performance for each bug report group separately.
We also investigate how the runtime performance of \bert, which utilizes fine grained matching, compares to other BERT-based architectures that rely on embedding aggregation and perform retrieval across the entire search space.
As baselines, we use (1) Locus~\cite{wen2016}, a state-of-the-art approach based on VSM that locates bug-inducing changesets, and (2) TBERT-Single and TBERT-Siamese~\cite{lin2021traceability} approaches that utilize aggregated BERT-based representations that have recently been proposed for software engineering. 

\smallskip
\noindent
\textbf{RQ2:} \textit{Which changeset encoding strategy is the most profitable? Are there advantages to using hunks, changeset-files or entire changesets as the primary data dimension?}

\noindent
In this RQ, we first investigate whether encoding information about the type of modification in each line of a changeset can increase the performance of the \bert model. We evaluate two alternatives to encode changesets semantics, \arc , \arcl, and a baseline approach, \de, which disregards change-related information.  
Second, we investigate how granularity of the input data affects the model performance and what are the benefits and challenges of leveraging changesets, changeset-files, or hunks in our model.
To answer this RQ, we fine tune \bert separately for each of the encoding strategies and with each input data granularity, resulting in 9 evaluation configurations per software project, measuring the model's performance in retrieving relevant changesets.

\subsection{Dataset and baselines}
To answer the RQs, we leverage the dataset of bugs and their inducing changesets collected and manually validated by Wen et al.~\cite{wen2016}; manually validated datasets remove the error that can be introduced by the SZZ algorithm that maps the bug fixing to the inducing commit~\cite{neto2018impact}. This dataset includes 6 software projects, namely AspectJ, JDT, PDE, SWT, Tomcat and ZXing (descriptive statistics are presented in Table~\ref{tab:dataset}).
To create a training set for each project, we selected the first half of project's pairs of bug reports and bug-inducing changesets, ordered by bug opening date, as a training set, and left the remaining half as a test set. 
For each pair in the training sets, we also create a negative sample by randomly choosing a code change which does not belong to the inducing changeset, essentially forming triplets of bug report, bug-inducing changeset, not bug-inducing changeset.
We experimented with choosing negative samples by selecting a syntactically similar changeset that was not bug-inducing but we did not observe a significant change in retrieval accuracy. As this type of generating negative samples incurred substantial computational cost to gather, we opted to use random sampling. 
Finally, for each project we obtained a balanced training set with equal number of positive and negative examples. 
Note that although training sets do not include all available code changes, during bug localization the model performs retrieval across \textit{all} code changes available for a specific project (as explained in Section \ref{sec:late-interaction}).
To study the impact of different changeset data granularity on the BERT-based models, we created a separate dataset for each type of granularity, i.e., changesets, changeset-files and hunks. To this end, for changeset-file and hunk granularity, we divide the bug-inducing changeset to file- or hunk-level code changes, such that one bug report creates multiple pairs with files or hunks from its respective inducing changeset.
\begin{table}[tb]
\centering
\footnotesize
\caption{Projects in evaluation dataset.}
\begin{tabular}{lrrrr}
\toprule
                 & \multicolumn{1}{c}{\textbf{\begin{tabular}[c]{@{}c@{}}\#Bugs\end{tabular}}} & \multicolumn{1}{c}{\textbf{\#Changesets}} & \multicolumn{1}{c}{\textbf{\#Changeset-files}} & \multicolumn{1}{c}{\textbf{\#Hunks}} \\ \midrule
\textbf{AspectJ} & 200     & 2,939    & 14,030  & 23,446   \\
\textbf{JDT}     & 94     & 13,860   & 58,619  & 150,630  \\
\textbf{PDE}     & 60     & 9,419    & 42,303  & 100,373  \\
\textbf{SWT}     & 90     & 10,206   & 25,666  & 69,833   \\
\textbf{Tomcat}  & 193    & 10,034   & 30,866  & 72,134   \\ 
\textbf{ZXing}   & 20     & 843      & 2,846   & 6,165  \\\bottomrule
\end{tabular}
\label{tab:dataset}

\end{table}

We compare the performance of the proposed model with Locus~\cite{wen2016}, which is an unsupervised model that utilizes hunk-level granularity and the VSM to locate relevant changesets based on the maximum similarity score obtained between a bug report, a hunk, and a log message.
Note that \bert does not use log messages as our goal is to explore mapping from natural language in a bug report to code changes. While well written log messages can have a positive impact on the results by boosting the scores for some changesets, not all relevant code changes are accompanied by logs of good quality~\cite{maalej2010can,jiang2017automatically}. 
As a second set of baselines, we employ TBERT architectures for software artifacts retrieval recently proposed by Lin et al.~\cite{lin2021traceability}. Out of the three architectures investigated by Lin et al., we selected TBERT-Single and  TBERT-Siamese as our baselines, rejecting TBERT-Twin, since its performance in terms of accuracy and time was significantly surpassed by the two others. In general, both of these architectures are fairly similar to those presented in Fig.~\ref{fig:berts} with an exception of using more advanced embedding aggregation operators~\cite{lin2021traceability}.

\subsection{Metrics}
To evaluate the performance of the model, we employ a set metrics commonly used to evaluate performance of IR systems.
\setlength{\abovedisplayskip}{3pt}
\setlength{\belowdisplayskip}{3pt}

\smallskip
\noindent
\textbf{Mean Reciprocal Rank:} MRR quantifies the ability of a model to locate the first relevant changeset to a bug report. The metric is calculated as an average of reciprocal ranks across $B$ bug reports, while a reciprocal rank for a bug report $B_i$ is equal to an inverted rank of the first relevant changeset in the ranking:
$$MRR = \frac{1}{|B|} \sum_{i=1}^{|B|} \frac{1}{1stRank_{B_i}}.$$

\noindent
\textbf{Mean Average Precision:} MAP measures how well a model can locate all changesets relevant to a bug report. MAP is calculated as the mean of average precision values ($AvgP$) for $B$ bug reports, while average precision for a bug report $B_i$, $AvgP_{B_i}$, is computed based on the positions of all relevant changesets in the ranking:
$$MAP = \frac{1}{|B|} \sum_{i=1}^{|B|} \frac{1}{AvgP_{B_i}}.$$

\noindent
\textbf{Precision@K:} P@K evaluates how many of the top-$K$ changeset in a ranking are relevant to a bug report. The value of P@K is equal to the number of relevant changesets $|Rel_{B_i}|$ located in the top-$K$ position in the ranking averaged across $B$ bug reports:
$$P@n = \frac{1}{|B|} \sum_{i=1}^{|B|} \frac{|Rel_{B_i}|}{K}.$$

\subsection{Experiment setup} \label{sec:experiment-seup}

The experiments were conducted on a server with Dual 12-core 3.2GHz Intel Xeon and utilized 1 NVIDIA Tesla V100 with 32GB RAM memory running on CUDA version 10.1. To implement our model, we used PyTorch v.1.7.1, HuggingFace library v.4.3.2, and Faiss v.1.6.5 with GPU support.
Since pre-training is a computationally expensive task and requires a huge dataset, we decided to use an available pre-trained BERT model, BERTOverflow~\cite{tabassum2020code}. BERTOverflow is trained on StackOverflow data, hence it contains a mixture of code snippets and natural language descriptions, which is logical for the bug localization task that operates on both code and natural language.
We fine tuned our BERT model and TBERT baselines for 4 epochs with batches of size 16 and a learning rate of 3E-06~\cite{devlin2019bert}. Based on the average number of tokens in bug reports, hunks, changeset-files and changesets across the evaluation projects, we set the maximum length limit to 256, 256, 512, and 512 respectively. All input documents are truncated or padded to their respective length limit.
For the Faiss index, we set the number of partitions to 320 and retrieved a total of 1000 changesets for re-ranking with \bert~\cite{khattab2020colbert}. In the case of Locus, we set the model parameters to $\lambda=5$ and $\beta_2=0.2$, indicated by the authors to provide the highest performance.

%% file: results.tex
\section{Results}

\subsection{RQ1: Retrieval performance}

\begin{table}[t]
\centering
\footnotesize
\caption{Mean Reciprocal Rank (MRR) of changeset-based BL techniques for different types of bug reports.}
\begin{tabular}{p{1.1cm}|l|rrr|r|r}
\toprule
& & \multicolumn{5}{c}{\textbf{Bug report type}} \\ \cmidrule(lr){3-7}
\textbf{Technique}  &  \textbf{Granularity} & \multicolumn{1}{p{0.6cm}}{BL\textsubscript{NL}\newline \scriptsize{\em n=151}} & \multicolumn{1}{p{0.6cm}}{BL\textsubscript{PL}\newline \scriptsize{\em n=75}} & \multicolumn{1}{p{0.6cm}}{BL\textsubscript{FL}\newline \scriptsize{\em n=105}} & \multicolumn{1}{p{0.7cm}}{BL\textsubscript{NL+PL}\newline \scriptsize{\em n=226}}& \multicolumn{1}{p{0.9cm}}{All BRs\newline \scriptsize{\em n=331}} \\ \midrule
 \midrule
\textit{Locus} & Hunks & 0.235 & 0.302 & \textbf{0.452} & 0.258 & 0.319 \\ \midrule
\textit{TBERT-} & Changesets & 0.119 & 0.213 & 0.136 & 0.150 & 0.146 \\
\textit{Single} & Change. files & 0.274 & 0.469 & 0.299 & 0.339 & 0.326 \\
 & Hunks & 0.268 & 0.429 & 0.273 & 0.321 & 0.306 \\ \midrule
\textit{TBERT-}  & Changesets & 0.125 & 0.256 & 0.080 & 0.168 & 0.140 \\
\textit{Siamese}  & Change. files & 0.263 & 0.424 & 0.200 & 0.316 & 0.279 \\
 & Hunks & 0.236 & 0.333 & 0.171 & 0.269 & 0.238 \\ \midrule
\textit{FBL-BERT} & Changesets & 0.076 & 0.114 & 0.113 & 0.089 & 0.096 \\
   & Change. files & \textbf{0.303} & 0.441 & 0.294 & 0.349 & 0.331 \\
 & Hunks & 0.290 & \textbf{0.509} & 0.338 & \textbf{0.363} & \textbf{0.355} \\ \bottomrule
\end{tabular}
 \label{tab:rq1}
\end{table}

\textbf{Retrieval accuracy.}
Table~\ref{tab:rq1} contrasts the retrieval performance of the \bert model against the baseline approaches for three different types of bug reports: not localized, partially localized, or fully localized. If a bug report has no mentions of relevant classes, it is classified as not localized (\nl); when some of the relevant classes appear in the report, the bug is categorized as partially localized (\pl); and if all relevant class names are provided, the bug report is fully localized (\fl)~\cite{kochhar2014}.
Note that in the case of \bert, we use the results of the model trained with {ARC\textsubscript{L}} encoding since, on average, it provides the best performance across the evaluation projects, as shown in Section~\ref{sec:rq2}. 

\bert outperforms Locus for \nl and \pl by 5.5\% and 20.6\% respectively,
while in the case of \fl, Locus surpasses our approach by 11.4\%. Given that Locus relies on more direct term matching between a bug report and a changeset, it makes intuitive sense that such a model performs best when localization hints are present in a bug report, and struggles in their absence (as indicated by lower MRR values for \nl and \pl). On the other hand, \bert utilizes higher-level association between bug reports and bug-introducing changesets, which can result in exact matches getting less emphasis. Interestingly, the highest improvement in retrieval accuracy is observed for \pl indicating that the model can effectively retrieve changesets based on partial clues by associating them with patterns learned from historical data. 

The performance of both TBERT models and \bert improves when the models are trained and evaluated on hunks or changeset-files. Compared to leveraging changesets, across all bug reports \bert improves between 23.5\%--25.9\%, while the retrieval accuracy of TBERT-Single and TBERT-Siamese increases by 16\%--18\% and 9.8\%--13.9\% respectively. 
While this results indicate that leveraging fine grained data affects retrieval performance positively, it is important to note that the poor performance observed for changesets can be partially attributed to the input size limit of the BERT model (i.e., 512 tokens), which is more often exceed by changesets than hunks or changeset-files. More specifically, in our dataset truncation affects about 8\% of hunks and 25\% of changeset-files compared to 45\% of changesets.

In general, \bert outperforms TBERT-Single and TBERT-Siamese by 4.9\% and 7.6\% respectively across all types of bug reports. Comparing the results of \bert trained on hunks to TBERT models trained on changeset-files, given that changeset-files provide on average the best performance for TBERT models, we note varying difference in retrieval accuracy depending on the bug report type.
In the case of \nl, \bert improves MRR score by only about 2\% over TBERT models. 
For \pl, \bert improves by 4\% and 8.5\% over TBERT-Single and TBERT-Siamese, while for \fl the improvement is equal to 3.9\% and 13.8\% respectively. The larger gap in retrieval accuracy for \pl and \fl between \bert and TBERT models indicates the importance of token-level embedding matching, i.e., while TBERT uses aggregated embedding to represent and compare documents, the token-level embedding matching performed by \bert allows this model to better recognize the key code names presented in the bug report, which, in turn, translates to higher retrieval accuracy.

\noindent
\textbf{Retrieval time.}
One of the key desirable characteristics of \bert is to perform efficient retrieval across a large corpus. This would allow it to leverage fine grained data, such as changesets-files or hunks which were observed to provide the best retrieval accuracy, while maintaining reasonable retrieval delay.
In Fig.~\ref{fig:performance}, we compare the average retrieval time per bug report with respect to the increasing number of documents in the search space, i.e., changesets, changesets-files and hunks.
In general, \bert retrieves relevant documents faster than both TBERT models with the retrieval time gap increasing as the search space grows. More specifically, TBERT-Single is the slowest model and requires about 50s to perform retrieval over a small number of documents (e.g., ZXing), and nearly 1000s(!) for a large project (e.g., JDT). TBERT-Siamese is significantly faster than TBERT-Single, and up to the search space of about 15K documents, it performs on-pair with \bert. However, after that point, retrieval time for TBERT-Siamese rises steadily to reach about 70s for the largest search space, while in the case of \bert the retrieval time is still just above 1s.
By comparing the performance of \bert against TBERT models, it becomes evident that plain BERT-based models can quickly hit a retrieval delay wall which makes them impractical to use. On the other hand, \bert scales up with respect to the search space size allowing to leverage fine grained data to increase retrieval accuracy without sacrificing model responsiveness.

Note that the observed speed improvement is the result of both \bert and FAISS. More specifically, the training objective of \bert (i.e., finding most similar embedding vectors) \textit{enables} using vector similarity search (e.g., FAISS). 
As a consequence, FAISS can be used to retrieve the $K$ best candidates ($K<<N$, where $N$ is \#documents) with similar word-level embedding representations that are then re-ranked by \bert. By re-ranking only $K$ documents, the search space becomes significantly reduced, hence decreasing the retrieval time.
On the other hand, typical BERT-based pipelines (e.g., TBERT) concatenate bug reports and changesets, and use neural network layers to estimate a relevancy score. This approach precludes pruning the search space via FAISS, therefore, during retrieval a bug report has to be compared to all $N$ documents, which in turn increases retrieval delay.

\definecolor{aspectj}{HTML}{7fc97f}
\definecolor{jdt}{HTML}{beaed4}
\definecolor{pde}{HTML}{fdc086}
\definecolor{swt}{HTML}{707070}
\definecolor{tomcat}{HTML}{f0027f}
\definecolor{zxing}{HTML}{386cb0}

\begin{figure}
\centering
\begin{tikzpicture}

  \begin{axis}
[
    scale only axis, 
    height=4cm,
    width=0.4\textwidth,
    title style={yshift=-1ex},
    xlabel={Number of documents},
    ylabel={Time [s]},
    xlabel style={font=\small},
    ylabel near ticks,
    ymode=log,
    y tick label style={
			font=\footnotesize,
	},
    x tick label style={
        /pgf/number format/fixed,
		font=\footnotesize,
	},
	ylabel style={font=\small},
	xticklabels={0,20k,40k,60k,80k,100k,120k,160k},
	xtick={0,20000,40000,60000,80000,100000,120000,160000},
	scaled x ticks = false,
	ymajorgrids=true,
	xmajorgrids=true,
	grid style={dashed,white!90!black},
    legend style={font=\scriptsize,column sep=1ex,draw opacity=0.8, at={(0.285,0.0)},anchor=south west},
    legend columns=3,
    legend entries={AspectJ,JDT,PDE,SWT,Tomcat,ZXing},
  ]
  
    \addlegendimage{mark=none, aspectj, solid, very thick};
	\addlegendimage{mark=none,jdt,solid, very thick};
	\addlegendimage{mark=none,pde,solid, very thick};
	\addlegendimage{mark=none,swt,solid, very thick};
	\addlegendimage{mark=none,tomcat,solid, very thick};
	\addlegendimage{mark=none,zxing,solid, very thick};
	
    \addplot[only marks, mark=*, draw=aspectj, mark options={fill=aspectj,scale=1.2}]
    coordinates{ 
      (2939,94.90219753073076)
    }; 
    \addplot[only marks, mark=*, draw=jdt, mark options={fill=jdt,scale=1.2}]
    coordinates{ 
      (13860,326.31497849443906)
    }; 
    \addplot[only marks, mark=*, draw=pde, mark options={fill=pde,scale=1.2}]
    coordinates{ 
      (9419,177.99710344446117)
    }; 
    \addplot[only marks, mark=*, draw=swt, mark options={fill=swt,scale=1.2}]
    coordinates{ 
      (10206,189.92723711495546)
    }; 
    \addplot[only marks, mark=*, draw=tomcat, mark options={fill=tomcat,scale=1.4}]
    coordinates{ 
      (10034,186.31348437443376)
    }; 
    \addplot[only marks, mark=*, draw=zxing, mark options={fill=zxing,scale=1.2}]
    coordinates{ 
      (843,0.2369999885559082)
    };

    \addplot[only marks, mark=*, draw=aspectj, mark options={fill=aspectj,scale=1.2}]
    coordinates{ 
      (14030,209.7668834959419)
    }; 
    \addplot[only marks, mark=*, draw=jdt, mark options={fill=jdt,scale=1.2}]
    coordinates{ 
      (58619,806.1636304337045)
    }; 
    \addplot[only marks, mark=*, draw=pde, mark options={fill=pde,scale=1.2}]
    coordinates{ 
      (42303,532.6981379328103)
    }; 
    \addplot[only marks, mark=*, draw=swt, mark options={fill=swt,scale=1.2}]
    coordinates{ 
      (25666,347.9953809522447)
    }; 
    \addplot[only marks, mark=*, draw=tomcat, mark options={fill=tomcat,scale=1.2}]
    coordinates{ 
      (30866,412.69782291601103)
    }; 
    \addplot[only marks, mark=*, draw=zxing, mark options={fill=zxing,scale=1.2}]
    coordinates{ 
      (2846,35.174333333969116)
    };

    \addplot[only marks, mark=*, draw=aspectj, mark options={fill=aspectj,scale=1.2}]
    coordinates{ 
      (23446,180.80396116590038)
    }; 
    \addplot[only marks, mark=*, draw=jdt, mark options={fill=jdt,scale=1.2}]
    coordinates{ 
      (150630,986.5645000001658)
    }; 
    \addplot[only marks, mark=*, draw=pde, mark options={fill=pde,scale=1.2}]
    coordinates{ 
      (100373,603.2270344865733)
    }; 
    \addplot[only marks, mark=*, draw=swt, mark options={fill=swt,scale=1.2}]
    coordinates{ 
      (69883,438.8665714263916)
    }; 
    \addplot[only marks, mark=*, draw=tomcat, mark options={fill=tomcat,scale=1.2}]
    coordinates{ 
      (72134,465.17076041549444)
    }; 
    \addplot[only marks, mark=*, draw=zxing, mark options={fill=zxing,scale=1.2}]
    coordinates{ 
      (6165,36.69777777459886)
    }; 
    
    \addplot[only marks, mark=triangle*, draw=aspectj, mark options={fill=aspectj,scale=1.4}]
    coordinates{ 
      (2939,0.8749805816168924)
    }; 
    \addplot[only marks, mark=triangle*, draw=jdt, mark options={fill=jdt,scale=1.4}]
    coordinates{ 
      (13860,1.3711304353631062)
    }; 
    \addplot[only marks, mark=triangle*, draw=pde, mark options={fill=pde,scale=1.4}]
    coordinates{ 
      (9419,0.9072758658178921)
    }; 
    \addplot[only marks, mark=triangle*, draw=swt, mark options={fill=swt,scale=1.4}]
    coordinates{ 
      (10206,2.822880954969497)
    }; 
    \addplot[only marks, mark=triangle*, draw=tomcat, mark options={fill=tomcat,scale=1.4}]
    coordinates{ 
      (10206,1.0322395811478298)
    }; 
    \addplot[only marks, mark=triangle*, draw=zxing, mark options={fill=zxing,scale=1.4}]
    coordinates{ 
      (843,0.35744444529215497)
    };

    \addplot[only marks, mark=triangle*, draw=aspectj, mark options={fill=aspectj,scale=1.4}]
    coordinates{ 
      (14030,1.4143203902013093)
    }; 
    \addplot[only marks, mark=triangle*, draw=jdt, mark options={fill=jdt,scale=1.4}]
    coordinates{ 
      (58619,10.898891303850257)
    }; 
    \addplot[only marks, mark=triangle*, draw=pde, mark options={fill=pde,scale=1.4}]
    coordinates{ 
      (42303,7.480103443408835)
    }; 
    \addplot[only marks, mark=triangle*, draw=swt, mark options={fill=swt,scale=1.4}]
    coordinates{ 
      (25666,2.6435952356883456)
    }; 
    \addplot[only marks, mark=triangle*, draw=tomcat, mark options={fill=tomcat,scale=1.4}]
    coordinates{ 
      (30866,5.1723437507947287)
    }; 
    \addplot[only marks, mark=triangle*, draw=zxing, mark options={fill=zxing,scale=1.4}]
    coordinates{ 
      (2846,0.294777790705363)
    };
    
    \addplot[only marks, mark=triangle*, draw=aspectj, mark options={fill=aspectj,scale=1.4}]
    coordinates{ 
      (23446,3.2972912603211633)
    }; 
    \addplot[only marks, mark=triangle*, draw=jdt, mark options={fill=jdt,scale=1.4}]
    coordinates{ 
      (150630,50.794434780659884)
    }; 
    \addplot[only marks, mark=triangle*, draw=pde, mark options={fill=pde,scale=1.4}]
    coordinates{ 
      (100373,22.060827584102237)
    }; 
    \addplot[only marks, mark=triangle*, draw=swt, mark options={fill=swt,scale=1.4}]
    coordinates{ 
      (69883,8.860952382995968)
    }; 
    \addplot[only marks, mark=triangle*, draw=tomcat, mark options={fill=tomcat,scale=1.4}]
    coordinates{ 
      (72134,9.983885416885217)
    }; 
    \addplot[only marks, mark=triangle*, draw=zxing, mark options={fill=zxing,scale=1.4}]
    coordinates{ 
      (6165,0.6214444372389052)
    };

    
    \addplot[only marks, mark=square*, draw=aspectj, mark options={fill=aspectj,scale=1.2}]
    coordinates{ 
      (2939,0.2239206276956152)
    }; 
    \addplot[only marks, mark=square*, draw=jdt, mark options={fill=jdt,scale=1.2}]
    coordinates{ 
      (13860,0.572050799714758)
    }; 
    \addplot[only marks, mark=square*, draw=pde, mark options={fill=pde,scale=1.2}]
    coordinates{ 
      (9419,0.5276943683624268)
    }; 
    \addplot[only marks, mark=square*, draw=swt, mark options={fill=swt,scale=1.2}]
    coordinates{ 
      (10206,0.4886723197236353)
    }; 
    \addplot[only marks, mark=square*, draw=tomcat, mark options={fill=tomcat,scale=1.2}]
    coordinates{ 
      (10034,0.4090411502462595)
    }; 
    \addplot[only marks, mark=square*, draw=zxing, mark options={fill=zxing,scale=1.2}]
    coordinates{ 
      (843,0.2318818926811218)
    }; 
    
    \addplot[only marks, mark=square*, draw=aspectj, mark options={fill=aspectj,scale=1.2}]
    coordinates{ 
      (14030,0.440366590609316)
    }; 
    \addplot[only marks, mark=square*, draw=jdt, mark options={fill=jdt,scale=1.2}]
    coordinates{ 
      (58619,1.1255723567719156)
    }; 
    \addplot[only marks, mark=square*, draw=pde, mark options={fill=pde,scale=1.2}]
    coordinates{ 
      (42303,0.9877877791722616)
    }; 
    \addplot[only marks, mark=square*, draw=swt, mark options={fill=swt,scale=1.2}]
    coordinates{ 
      (25666,0.6618046663245376)
    }; 
    \addplot[only marks, mark=square*, draw=tomcat, mark options={fill=tomcat,scale=1.2}]
    coordinates{ 
      (30866,0.6218135208663546)
    }; 
    \addplot[only marks, mark=square*, draw=zxing, mark options={fill=zxing,scale=1.2}]
    coordinates{ 
      (2846,0.6582082629203796)
    };

    \addplot[only marks, mark=square*, draw=aspectj, mark options={fill=aspectj,scale=1.2}]
    coordinates{ 
      (23446,0.419357883148506)
    }; 
    \addplot[only marks, mark=square*, draw=jdt, mark options={fill=jdt,scale=1.2}]
    coordinates{ 
      (150630,1.5801971871802147)
    }; 
    \addplot[only marks, mark=square*, draw=pde, mark options={fill=pde,scale=1.2}]
    coordinates{ 
      (100373,1.211761740843455)
    }; 
    \addplot[only marks, mark=square*, draw=swt, mark options={fill=swt,scale=1.2}]
    coordinates{ 
      (69883,0.8427713428224836)
    }; 
    \addplot[only marks, mark=square*, draw=tomcat, mark options={fill=tomcat,scale=1.2}]
    coordinates{ 
      (72134,0.7367574884483852)
    }; 
    \addplot[only marks, mark=square*, draw=zxing, mark options={fill=zxing,scale=1.2}]
    coordinates{ 
      (6165,0.7671671271324157)
    }; 
    
    \addplot [dashed, no markers, domain=1000:160000, samples=1000] {(0.0127*x)+37.252};
    \addplot [dashed, no markers, domain=5000:160000, samples=1000] {(0.0003*x)-3.3685};
    \addplot [dashed, no markers, domain=1000:160000, samples=1000] {(8e-06*x)+0.2959};
    \end{axis}

\end{tikzpicture}
\caption[Average retrieval time per a bug report with different sizes of search space.]{Average retrieval time per a bug report with different sizes of search space (\mycircle{gray2} TBERT-Single, \mytriangle{gray2} TBERT-Siamese, \mysquare{gray2}~\bert).}
\label{fig:performance}
\vspace{-2mm}
\end{figure}
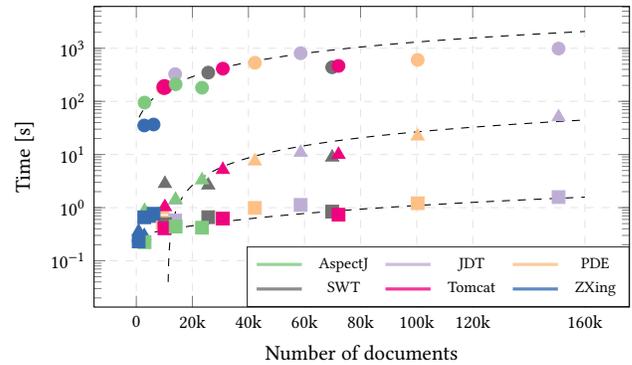

\definecolor{highlight}{HTML}{78c679}
\definecolor{highlight2}{HTML}{c2e699}
\definecolor{highlight3}{HTML}{ffffcc}
\begin{table*}[ht]
\centering
\footnotesize
\caption{Retrieval performance for different configurations of \bert.}
\begin{tabular}{l|l|lllll|lllll|lllll} 
\toprule
& \#Bugs &  \multicolumn{1}{c}{MRR}   & MAP        & P@1   & P@3    & P@5     & MRR   & MAP   & P@1   & P@3   & P@5   & MRR  & MAP   & P@1   & P@3   & P@5  \\ \midrule \midrule
\multicolumn{2}{l|}{\textbf{Changesets}}       & \multicolumn{5}{c|}{D}  & \multicolumn{5}{c|}{ARC} & \multicolumn{5}{c}{ARC\textsubscript{L}} \\ \midrule
        
\quad AspectJ & 104 & 0.053                     & 0.032      & 0.029 & 0.024  & 0.037   & 0.107 & 0.061 & 0.058 & 0.080 & 0.083 & 0.070 &0.042 & 0.029 & 0.045 & 0.044 \\
\quad JDT     & 47 & 0.097                     & 0.014      & 0.043 & 0.028  & 0.021   & 0.118 & 0.160 & 0.064 & 0.043 & 0.030 & 0.118 & 0.016 & 0.064 & 0.035 & 0.026 \\
\quad PDE     & 60 & 0.091                     & 0.012      & 0.067 & 0.022  & 0.020   & 0.099 & 0.019 & 0.033 & 0.033 & 0.031 & 0.103 & 0.013 & 0.067 & 0.033 & 0.027 \\
\quad SWT     & 43 & 0.067                     & 0.015      & 0.023 & 0.027  & 0.026   & 0.033 & 0.006 & 0.023 & 0.008 & 0.005 & 0.018 & 0.007 & 0.0   & 0.0   & 0.0   \\
\quad Tomcat  & 97 & 0.135                     & 0.048      & 0.052 & 0.070  & 0.074   & 0.132 & 0.051 & 0.062 & 0.072 & 0.071 & 0.141 & 0.055 & 0.062 & 0.077 & 0.088 \\
\quad ZXing   & 10 & 0.127                     & 0.034      & 0.100 & 0.033  & 0.020   & 0.141 & 0.034 & 0.100 & 0.033 & 0.040 & 0.155 & 0.061 & 0.100 & 0.133 & 0.120 \\ \midrule 
\quad \textit{All projects}  & \textit{331} & \textit{0.091} & \textit{0.030}     & \textit{0.042} & \textit{0.039}  & \textit{0.042}   & \textit{0.107} & \textit{0.040} & \textit{0.054} & \textit{0.057} & \textit{0.056} & \textit{0.096} & \textit{0.036} & \textit{0.045} & \textit{0.049} & \textit{0.049} \\ \midrule
\midrule

\multicolumn{2}{l|}{\textbf{Changeset-files}}        & \multicolumn{5}{c|}{D}  & \multicolumn{5}{c|}{ARC} & \multicolumn{5}{c}{ARC\textsubscript{L}} \\ \midrule
\quad AspectJ & 104 & 0.173                     & 0.083      & 0.154 & 0.085  & 0.100   & 0.165 & 0.079 & 0.144 & 0.085 & 0.085 & \cellcolor{highlight2}0.176 & \cellcolor{highlight2}0.085 & \cellcolor{highlight2}0.154 & \cellcolor{highlight2}0.095 & \cellcolor{highlight2}0.097 \\
\quad JDT    &  47 & \cellcolor{highlight2}0.403                     & \cellcolor{highlight2}0.060      & \cellcolor{highlight2}0.319 & \cellcolor{highlight2}0.184  & \cellcolor{highlight2}0.128   & 0.355 & 0.060 & 0.255 & 0.149 & 0.126 & \cellcolor{highlight3}0.368 & \cellcolor{highlight3}0.055 & \cellcolor{highlight3}0.277 & \cellcolor{highlight3}0.149 & \cellcolor{highlight3}0.109 \\
\quad PDE    & 30  & \cellcolor{highlight3}0.259                     & \cellcolor{highlight3}0.087      & \cellcolor{highlight3}0.167 & \cellcolor{highlight3}0.128  & \cellcolor{highlight3}0.101   & 0.236 & 0.069 & 0.133 & 0.117 & 0.094 & \cellcolor{highlight2}0.260 & \cellcolor{highlight2}0.079 & \cellcolor{highlight2}0.167 & \cellcolor{highlight2}0.128 & \cellcolor{highlight2}0.151 \\
\quad  \textbf{SWT} & 43   & \cellcolor{highlight2}0.552                     & \cellcolor{highlight2}0.129      & \cellcolor{highlight2}0.535 & \cellcolor{highlight2}0.217  & \cellcolor{highlight2}0.164   & \cellcolor{highlight3}0.538 & \cellcolor{highlight3}0.127 & \cellcolor{highlight3}0.535 & \cellcolor{highlight3}0.209 & \cellcolor{highlight3}0.159 & \cellcolor{highlight}0.555 & \cellcolor{highlight}0.131 & \cellcolor{highlight}0.535 & \cellcolor{highlight}0.233 & \cellcolor{highlight}0.173 \\
\quad Tomcat & 97 & 0.424                     & 0.099      & 0.361 & 0.175  & 0.147   & 0.421 & 0.116 & 0.351 & 0.191 & 0.155 & 0.\cellcolor{highlight2}463 & \cellcolor{highlight2}0.114 & \cellcolor{highlight2}0.381 & \cellcolor{highlight2}0.222 & \cellcolor{highlight2}0.183 \\
\quad ZXing & 10 & 0.199                     & 0.157      & 0.100 & 0.133  & 0.140   & 0.212 & 0.163 & 0.100 & 0.133 & 0.220 & 0.200 & 0.159 & 0.100 & 0.133 & 0.120 \\ \midrule
\quad \textit{All projects}  & \textit{331} & \textit{0.348}  & \textit{0.097}     & \textit{0.293} & \textit{0.162}  & \textit{0.138}   & \textit{0.325} & \textit{0.095} & \textit{0.269} & \textit{0.149} & \textit{0.128} & \textit{0.331} & \textit{0.092} & \textit{0.281} & \textit{0.145} & \textit{0.127} \\
\midrule \midrule

\multicolumn{2}{l|}{\textbf{Hunks}}       & \multicolumn{5}{c|}{D}  & \multicolumn{5}{c|}{ARC} & \multicolumn{5}{c}{ARC\textsubscript{L}} \\  \midrule

\quad \textbf{AspectJ} & 104 & 0.175                     & 0.084      & 0.163 & 0.091  & 0.093   & \cellcolor{highlight3}0.176 & \cellcolor{highlight3}0.082 & \cellcolor{highlight3}0.163 & \cellcolor{highlight3}0.093 & \cellcolor{highlight3}0.083 & \cellcolor{highlight}0.183 & \cellcolor{highlight}0.093 & \cellcolor{highlight}0.173 & \cellcolor{highlight}0.111 & \cellcolor{highlight}0.099 \\
\quad \textbf{JDT}   &  47 & 0.362                     & 0.059      & 0.255 &0.135  & 0.122   & 0.322 & 0.049 & 0.213 & 0.149 & 0.109 & \cellcolor{highlight}0.429 & \cellcolor{highlight}0.062 & \cellcolor{highlight}0.319 & \cellcolor{highlight}0.195 & \cellcolor{highlight}0.167 \\
\quad \textbf{PDE}  & 30  & 0.249                     & 0.088      & 0.167 & 0.122  & 0.141   & \cellcolor{highlight}0.288 & \cellcolor{highlight}0.093 & \cellcolor{highlight}0.200 & \cellcolor{highlight}0.144 & \cellcolor{highlight}0.127 & 0.200 & 0.068 & 0.133 & 0.078 & 0.087 \\
\quad SWT &  43  & 0.510                     & 0.117      & 0.465 & 0.225  & 0.196   & 0.519 & 0.142 & 0.442 & 0.240  & 0.201  & 0.526 & 0.131 & 0.488 & 0.217 & 0.164 \\
 \quad \textbf{Tomcat} & 97 & 0.426                     & 0.135      & 0.289 & 0.211  & 0.191   & \cellcolor{highlight3}0.441 & \cellcolor{highlight3}0.140 & \cellcolor{highlight3}0.351 & \cellcolor{highlight3}0.211 & \cellcolor{highlight3}0.211 & \cellcolor{highlight}0.482 & \cellcolor{highlight}0.129 & \cellcolor{highlight}0.412 & \cellcolor{highlight} 0.216 & \cellcolor{highlight}0.182 \\
\quad \textbf{ZXing} & 10 & \cellcolor{highlight}0.334                     & \cellcolor{highlight}0.225      & \cellcolor{highlight}0.200 & \cellcolor{highlight}0.283  & \cellcolor{highlight}0.370   & \cellcolor{highlight3}0.306 & \cellcolor{highlight3}0.193 & \cellcolor{highlight3}0.200 & \cellcolor{highlight3}0.283 & \cellcolor{highlight3}0.270 & \cellcolor{highlight2}0.328 & \cellcolor{highlight2}0.210 & \cellcolor{highlight2}0.200 & \cellcolor{highlight2}0.233 & \cellcolor{highlight2}0.240 \\ \midrule
\quad \textit{All projects}  & \textit{331} & \textit{0.330} & \textit{0.101}     & \textit{0.254} & \textit{0.159}  & \textit{0.152}   & \textit{0.334} & \textit{0.105} & \textit{0.272} & \textit{0.162} & \textit{0.144} & \textit{0.355} & \textit{0.107} & \textit{0.296} & \textit{0.171} & \textit{0.149} \\ \midrule\bottomrule

\end{tabular}
\label{tab:rq2}
\end{table*}

\noindent
\textbf{Error analysis.}
To gain more insight into factors that negatively affect the retrieval accuracy of \bert, we manually analyzed the bug reports for which the model struggles the most. More specifically, we selected all bug reports where the bug-inducing hunk was ranked 50 or worse by \bert. This resulted in 20 bug reports ($BR_{NL}=8$, $BR_{PL}=3$, $BR_{FL}=9$) that the authors independently analyzed, contrasting the retrieved hunks to the true bug-inducing hunks in order to devise a set of common issues causing low retrieval accuracy. The authors also examined the most similar terms (and their weights) for both the retrieved and gold set hunks, focusing specifically on the sources of largest differences between the two. Finally, the authors discussed their independent observations and agreed on three common error categories: {\em stack trace/code snippets}, {\em comments}, and {\em code tokens splitting}, where a single bug report can belong to more than one error category. We discuss each of these, in turn. 

In 11 out of 20 bug reports, the difficulty to retrieve the correct hunk was caused by the presence of a code snippet or a stack trace in the bug report. Since code snippets and stack traces typically consists of multiple class names or code tokens, they have a potential to introduce noise through unrelated code names, which, in turn, can lead the model astray~\cite{wang15usefulness}. 
For 7 out of 20 bug reports, we noted that the model was misguided by source code comments present in the top-1 retrieved hunk. Since source code comments are formulated in natural language, a highly-contextual model like BERT tends to emphasize their similarity with the bug report as it is also expressed in natural language. 
For both of the above error categories, we believe that the wholesale removal of the problematic text (i.e, comments from code and code snippets and stack traces from bug reports) would negatively affect the model as it removes both relevant and irrelevant information. Hence, researchers should explore strategies to treat this data separately, perhaps by encoding their content within BERT with special tokens akin to the \arc and \arcl strategies we discuss in this paper.

Finally, for 5 of the bug reports, \bert failed due to spurious matches in code tokens that were split into sub-tokens during preprocessing. One of the previously observed strengths of BERT is in using the WordPiece algorithm to avoid the out-of-vocabulary problem by splitting unseen tokens into the largest sub-tokens that are part of the BERT vocabulary~\cite{karampatsis20oov}. Since source code identifier names are typically project-specific words, they do not occur in the pre-trained vocabulary, hence they are often split by WordPiece (e.g., ManagerServlet $\rightarrow$ manager, \#\#servlet). The sub-tokens can then spuriously match other terms, including sub-tokens from other split identifiers, but not the whole, unsplit term. Researchers in the biomedical domain recognized the same issue affecting medical terms and proposed domain-specific BERT adaptations~\cite{lee2019biobert, beltagy2019SciBERT, gu2021domainspecific}.

\subsection{RQ2: Changeset encoding strategy} \label{sec:rq2}
\smallskip

Table~\ref{tab:rq2} shows retrieval performance of \bert trained and evaluated with different changeset encoding strategies and input data granularities. For each project, the three best performing configurations  are highlighted, such that dark green marks a configuration with the highest retrieval performance, while green and yellow correspond to the second and third best configurations. Overall, we notice that using entire changesets as the granularity of input results in, by far, the worst performance across all of the investigated configurations for all evaluation projects. We can attribute this result to: (1) truncation of changesets due to input length limitation of the BERT model; and (2) tangled changes within a single changeset~\cite{herzig2013tangled}, which are likely to affect the model by introducing noise via unrelated code modifications. On the other hand, while the model based on hunks or changeset-files is not free of these problems, the finer data granularity allows it to partially overcome them. For instance, in case of tangled changes, dividing the entire changeset into hunks or changeset-files creates multiple new data points, which limits the noise introduced by instances that are poorly related to the bug. 
The difference in retrieval accuracy across all the metrics between using hunks and changeset-files as the input data is minor and differs from 1\% to 12.2\% per project. This results is indicative of the observation that leveraging hunks and changeset-files perform similarly and are both resilient to the problems affecting changesets. 

Examining the results for different changest encoding strategies, we observe that \arcl performs universally best across hunks and changeset-files. 
Interestingly, at the level of changeset-files, the baseline encoding \de, which does not encode modification type, does surprisingly well and outperforms \arc encoding. We attribute this result to the specifics of \arc encoding, which groups lines based on the performed modification, hence in the case of larger documents the grouping may affect the semantics of the documents. On the other hand, \arc encoding for hunks is less likely to be susceptible to that problem since hunks are typically much shorter. 
Analyzing the results for different projects, we observe that \arcl performs best for AspectJ, JDT and Tomcat, with an improvement in MRR scores of 0.7\%, 6.7\% and 4.1\% over their second best configurations respectively, while \arc is the most beneficial strategy for the PDE project. 
In the case of SWT, we observe the highest retrieval accuracy with \arcl, while ZXing performs best with \de encoding; however, both of these observations are likely negligible given the low difference between \arcl and other encodings for SWT, and the relatively fewer bug reports in the ZXing project.
Overall, we conclude that leveraging changesets semantics via encoding modification with either \arc and \arcl increases retrieval accuracy over the \de configuration which does not provide the model with additional information about the change. However, based on these results, the difference between \arc and \arcl is not significant enough to clearly indicate which strategy is superior on average.

\subsection{Threats to validity}
The conclusions of this paper suffer from several threats to validity. A key threat to the internal validity of our study are the specific parameter choices we used to build our \bert model. A mitigating factor is that all parameters were either studied by us or were reported in other prior reputable papers as recommended or optimal~\cite{devlin2019bert,khattab2020colbert}. Another threat is our automated separation of bug reports based on localization hints into, not localized, partially localized, and fully localized, which may result in mistaken categorization, even though we used a well-known and frequently followed procedure~\cite{kochhar2014}.

Leveraging changesets for bug localization poses another threat due to possible noise that can be introduced by SZZ~\cite{rosa2021evaluating}, which could result in poor quality mapping between bug reports and bug-inducing changesets. However, the dataset was validated manually~\cite{wen2016,zhou2012where}, and therefore such mistakes, if they still exist, should not significantly affect our conclusions. Errors due to tangled changes \cite{mcintosh2011build,herzig2013tangled} are still possible in the dataset as such changes are difficult to remove manually. We believe tangled commits to have affected our final presented results (as discussed in RQ2), however, since tangled commits are a part of software development removing them completely may arguably result in unrealistic evaluation.


A threat to external validity, which concerns the ability to generalize our evaluation results, is that we applied the bug localization technique only on a limited number of bugs collected from a selection of popular open source Java projects. A mitigating factor is that the projects have a variety of purposes and development styles and the benchmark we used has also been applied to prior changeset-based bug localization studies~\cite{saha_bluir_2013,wong_brtracer_2014,wen2016}.
Another threat to external validity is in the chosen evaluation metrics, which may not directly gauge user satisfaction with our bug localization technique~\cite{wang2015usefulness}, impacting the validity of the reported results. The threat is mitigated by the fact that the selected metrics are well-known and widely accepted as best available to measure and compare the performance of IR techniques.

%% file: related.tex
\section{Related work}
Bug localization has generated significant research interest over the years. In this section, first, we survey related code element-based bug localization techniques, followed by approaches towards bug-inducing changeset retrieval.
Finally, we review methods for encoding changesets characteristics.

\subsection{Code element-based bug localization}
Bug localization techniques predominantly utilize information retrieval where the bug report text is used to formulate a query that is matched to a corpus of code elements, i.e., classes or methods. 
To compute similarity between bug reports and source code, the Vector Space Model (VSM) is often used as one of the simplest and effective information retrieval algorithms, which is leveraged by many bug localization techniques. For instance, BugLocator~\cite{kim_buglocator_2013}
combines two rankings, one produced by similarity between the bug report and code elements and another based on similarity of the bug report to prior
fixed bug reports. BLUiR~\cite{saha_bluir_2013} uses code and bug report structure to create groups of terms and computes similarity between different groups separately, while AmaLgam~\cite{wang_amalgam_2014} creates an ensemble consisting of BugLocator, BLUiR and a defect predictor leveraging the development history of a project. BRTracer~\cite{wong_brtracer_2014} focuses on analyzing and prioritizing stack traces when they are included in bug reports. Kochhar et al. were among the first to report that evaluation of bug localization was biased by explicit localization hints in a significant subset of the included bug reports~\cite{kochhar2014}. VSM-based techniques are likely to perform well on such bug reports, though localizing them may not be as useful to developers~\cite{tabassum-etal-2020-code}. Mills et al. refute the idea that VSM-based bug localization are significantly aided by hints, and note that VSM can perform well for bug localization if more attention is paid to how the query is constructed from the bug report text~\cite{mills2020relationship}. However, their findings do not preclude additional accuracy improvements by using more complex, semantic models, such as BERT. 

More recently, software engineering researchers have been interested in the applying deep learning techniques towards bug localization. For instance, TRANP-CNN~\cite{deeptransfer} is a recent technique that combines cross-project transfer learning and convolutional neural networks to achieve state-of-the-art performance on file-level bug localization. 
CooBa improves on TRANP-CNN by combining a shared encoder to capture cross-project with per-project features and using adversarial training to ensure that the per-project information remains unaffected by noise~\cite{Zhu2020CooBaCB}. Lam et al.'s technique, DNNLOC, combines a deep neural network with the VSM in order to be effective across different types of similarity~\cite{lam_dnnloc}. While we also leverage a deep learning model, BERT is significantly different from these prior techniques. Recent work in bug localization also includes reports on the value of retrieving changesets instead of source code elements~\cite{murali2020industry,wen2016}.

\subsection{Changeset-based bug localization}
The earliest work on changeset-based bug localization is Locus~\cite{wen2016}, which is based on VSM matching of bug reports to hunks. To adjust for localization hints, Locus adapts its similarity scores based on the proportion of code element mentions in a bug report. Bhagwan et al.~\cite{bhagwan2018orca} introduced Orca, a tool that uses a  provenance graph to identify commits leading to faulty builds. 
ChangeLocator~\cite{wu2018change} uses historical data on software crashes to build a model identifying relevant changesets based on collection on crash reports. Although this approach allows to retrieve changesets, it requires sufficient amount of historical data to train the model, and a stack trace as an input. One of the benefits of VSM is that it is an unsupervised approach, hence a training corpus of bug reports and their inducing commits is not required. However, as VSM fundamentally requires at least partial token overlap, while it ignores the context in which tokens appear in documents, all bug localization technique based on it have a limited accuracy ceiling~\cite{murali2020industry}.

Recently, researchers have also shifted their attention to deep learning models for changeset-based localization. For instance, Murali et al.~\cite{murali2020industry} proposed Bug2Commit, an unsupervised model leveraging multiple dimension of data associated with bug reports and commits, such as metrics, stack traces or commit meta data. They observed that using embeddings can lead to improvement in model accuracy when compared to BM25. Lin et al.~\cite{lin2021traceability} studied the trade-offs between different BERT architectures for the purpose of changeset retrieval, and observed the accuracy of Siamese architecture is on pair with Single-BERT architecture, while being significantly faster. However, the speed and interactivity of these models is not on par with the BERT technique described in this paper.

\subsection{Changeset representation}
Building a semantically rich representations of changesets is relevant to other software engineering applications beyond bug localization, i.e., just-in-time defect prediction, recommendation of a code reviewer for a patch, tangled change prediction.
Approaches that define novel changeset embeddings (vector representations of changeset), including CC2Vec~\cite{hoang2020cc2vec} and Commit2Vec~\cite{lozoya2019commit2vec}, leverage the difference between added and removed lines of code, among other changeset characteristics. Corley et al.~\cite{corley2018} studied how including different types of lines from a changeset affects the performance of Latent Dirichlet Allocation-based feature location, observing that including context, additions, and log messages, but excluding removed lines, achieves the best performance. However, these studies did not utilize a transfer learning technique, like BERT, which requires compatibility with a pre-trained model, and also prior work did not extensively explore hunks as a primary data dimension.

%% file: conclusion.tex
\section{Conclusion}
This paper presents an approach for automatically retrieving bug-inducing changesets for a newly reported bug. The approach uses the popular BERT model to more accurately match the semantics in the bug report text to the inducing changeset. More specifically, we describe the \bert model, based on the prior work by Khattab et al.~\cite{khattab2020colbert}, which speeds up the retrieval of results while performing fine grained matching across all embeddings in the two documents.
The results show an improvement in retrieval accuracy for bug reports that lack localization hints or have only partial hints. We also evaluate different approaches for utilizing changesets in BERT-like models, producing recommendations on the input data granularity and the use of special tokens for the purpose of capturing changeset semantics.